\pdfoutput=1

\documentclass[11pt]{article}

\usepackage{acl}

\usepackage{times}
\usepackage{latexsym}
\usepackage{amsmath,amssymb}  
\usepackage[mathscr]{eucal}
\usepackage{graphicx}
\usepackage{enumitem}

\usepackage[T1]{fontenc}

\usepackage[utf8]{inputenc}

\usepackage{microtype}

\title{Contrastive language and vision learning of
general fashion concepts }

\author{Patrick John Chia\thanks{* FashionCLIP was started by JT and FB. PC and GA led implementation and experiments; DG and ARM prepared the dataset, performed EDA and provided domain knowledge; ST and CG helped with fine-tuning, model evaluation and research background. Everybody contributed to the final draft. JT and FB acted as senior PIs for the project.} \\
  Coveo \\ Montreal, Canada \\
  \texttt{pchia@coveo.com} \\\And
  Giuseppe Attanasio \\ Bocconi University \\ Milan, Italy  \And Federico Bianchi \\
  Bocconi University \\ Milan, Italy \\\AND
  Silvia Terragni \\
  Telepathy Labs, Zurich \\\And
  Ana Rita Magalhães \\ Farfetch \\ Porto, Portugal \And Diogo Goncalves 
  \\ Farfetch \\ Porto, Portugal\\\AND
  Ciro Greco \\ Coveo Labs \\ New York, United States  \And Jacopo Tagliabue \\
  Coveo Labs \\ New York, United States \\
  }

\begin{document}

\onecolumn
\section*{Note by the authors (Apr. 18)}
This work reflected the ongoing research work on the topic at our organizations, and presented the first encouraging results we obtained at the end of 2021. 

In 2022, we were fortunate enough to iterate significantly on the original idea, and we ended up publishing a peer-reviewed article on Nat. Sci. Rep. titled \textbf{Contrastive language and vision learning of general fashion concepts}: \url{https://www.nature.com/articles/s41598-022-23052-9}. While \textit{this} draft is maintained on arxiv (with now the same title to avoid confusion), we encourage practitioners to read and cite our most recent publication, as it contains a better argued presentation on the topic of large and general models in industry, as well as more experiments and benchmarks.  

\begin{itemize}
    \item \textbf{Preferred citation}: \textit{Chia, P.J., Attanasio, G., Bianchi, F. et al. Contrastive language and vision learning of general fashion concepts. Sci Rep 12, 18958 (2022). https://doi.org/10.1038/s41598-022-23052-9}.
    \item \textbf{HuggingFace model}: \url{https://huggingface.co/patrickjohncyh/fashion-clip}.
    \item \textbf{GitHub Repository}: \url{https://github.com/patrickjohncyh/fashion-clip}.
\end{itemize}

\maketitle
\begin{abstract}
The steady rise of online shopping goes hand in hand with the development of increasingly complex ML and NLP models. While most use cases are cast as specialized supervised learning problems, we argue that practitioners would greatly benefit from more transferable representations of products. In \textit{this} work, we build on recent developments in contrastive learning to train \textit{FashionCLIP}, a \textit{CLIP}-like model for the fashion industry. We showcase its capabilities for retrieval, classification and grounding, and release our model and code to the community. 
\end{abstract}

\section{Introduction}

\begin{quote}
``Sê plural como o universo!''\footnote{``Be plural, like the universe!''.} -- F. Pessoa.
\end{quote}

The	extraordinary growth of online retail - as of 2020, 4 trillion dollars per year~\cite{emarketer2020} - had a profound impact on the fashion	industry, with 1 out of 4 transactions now happening online~\cite{McKinsey2020}. The combination of large amounts of data and variety of use cases supported by growing investments has made e-commerce fertile for the application of cutting-edge machine learning models, with NLP involved in  recommendations~\cite{Moreira2019OnTI,guo-etal-2020-deep,Goncalves2021}, information retrieval (IR)~\cite{10.1145/3459637.3482276}, 
product classification \cite{chen-etal-2021-multimodal-item} and many other use cases~\cite{Tsagkias2020ChallengesAR}. 

As a standard practice before the rise of capable zero-shot alternatives, e-commerce models are typically trained over task-specific datasets, directly optimizing for individual metrics: for example, a product classification model might be trained on $<product\,description, category>$ pairs derived from catalog data~\cite{Gupta2016ProductCI}. Inspired by~\citet{Radford2021LearningTV}, we leverage the connection between text and images in a large fashion catalog to learn generic,  multi-modal product concepts, and apply the resulting model to important language-related tasks. In particular, we summarize our contributions as follows:

\begin{enumerate}
    \item to the best of our knowledge, \textbf{\texttt{FashionCLIP}} is the first industry-specific \texttt{CLIP}-like model. The model is trained on over 700k ${<image,text>}$ pairs from the inventory of \textit{Farfetch}, one of the largest fashion luxury retailer in the world, and is applied to use cases known to be crucial in a vast global market;
    \item we evaluate \texttt{FashionCLIP} in a variety of tasks, showing that fine-tuning helps capture domain-specific concepts and generalize them in zero-shot scenarios; we supplement quantitative tests with qualitative analyses, and offer preliminary insights of how concepts grounded in a visual space unlock linguistic generalization;
    \item we transparently report training time, costs and emissions, and release to the community, under an open-source license, training code and plug-and-play checkpoints to help practitioners replicate and leverage our findings, while facilitating ROI considerations.
\end{enumerate}

We believe our methods and results are useful not just for the fashion industry, but broadly applicable to the ever-expanding universe of online retail. Aside from its practical significance, the evaluation in Section~\ref{sec:grounding} is new in the context of \texttt{CLIP}-like models, and we believe it may be of independent interest for future NLP work. 

\section{An industry perspective}
\label{sec:industry}

\begin{figure*}
  \centering
  \includegraphics[width=\textwidth]{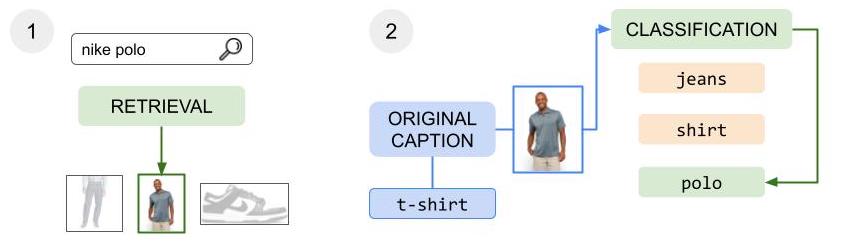}
  \caption{\textbf{Two important use cases in e-commerce}. (1) represents product retrieval through natural language; (2) represents an example of product classification using textual labels that differ from the original taxonomy: our hypothesis is that \texttt{FashionCLIP} concepts are transferable enough to make this zero-shot classification possible.}
  \label{fig:usecases}
  \vspace{-4mm}
\end{figure*}

By training \texttt{FashionCLIP} on \textit{Farfetch} inventory, we obtain transferable product representations that can be used on other shops. Our findings appeal to two kinds of practitioners: those working in e-commerce businesses with ever increasing inventory (e.g. \textit{Farfetch} itself), and those working in B2B technology providers which power \textit{hundreds} of e-commerce websites at the same time\footnote{To provide a frame of reference for this blooming market, Algolia, Lucidworks, Bloomreach recently raised more than USD100M  each~\cite{AlgoliaRound,LWRound,BloomreachRound}, and Coveo raised more than CAD 200M during its IPO in 2021 \cite{CoveoRound}.}. 
To evaluate the potential impact of \textit{this} work for the industry, we describe two core use cases where multi-modal concepts bring significant improvement (Fig.~\ref{fig:usecases}).

\textbf{\textit{Product search}} is one of the main channels of interactions and revenues between a shop and its users: 30\% to 60\% of total online revenues~\cite{bigCommerce2020,retailDive2020} are believed to come through product search. It is easy to see that at the scale of billion dollar marketplaces, such as \textit{Farfetch}, even marginal search improvements translate into sizable financial gains. 

Historically, product search has been performed mostly with textual features, by first matching queries and product descriptions in an index~\cite{tfidf,gillick2018end,Izacard2021TowardsUD} and then re-ranking the candidate results~\cite{10.1145/3219819.3219846}. However, there are good reasons to believe that including visual features can bring significant improvements, since images are often the most curated aspect of the catalog, whereas text quality varies throughout verticals, languages and specific product feeds. In addition, adding images can alleviate the negative impact of linguistic mismatch, for instance lexical differences between search queries (e.g. ``long-sleeved black blouse'') and marketing-oriented text in the catalog (e.g. ``the must-have touch for strong, independent women'') -- all particularly challenging in fashion, an industry with high customer expectations and many luxury items. \texttt{FashionCLIP} shows strong multi-modal retrieval performances even over unseen products (Section~\ref{sec:retrieval}), and provides relevance signals that complement incomplete or ambiguous descriptions (Appendix~\ref{sec:appendixOutput}). 

Our second main use case is \textbf{\textit{product classification}} -- i.e. predicting a product category given its meta-data. Classification is usually cast as a supervised learning problem, where golden labels are obtained from the catalog itself or collected through crowd-sourcing \cite{chen-etal-2021-multimodal-item,chen-miyake-2021-label}. Finding a good solution is pressing especially for multi-brand retailers and marketplaces, with fast growing catalogs and limited control, since categorization and textual data may vary wildly across different product suppliers. \texttt{FashionCLIP} alleviates the problem of data quality by providing classification without additional training: its zero-shot capabilities allow for quick classification of products in target classes of interest, \textit{irrespective of the specific labeling schemes of individual suppliers} (Section~\ref{sec:classification}). Transferable concepts help with the interoperability of overlapping and yet different fashion taxonomies \cite{interop}, a challenge increasingly recognized as central by both practitioners and commentators ~\cite{vogue}\footnote{This includes the case of catalogs in low-resources languages, for which an English classification is still desirable.}. This would not be possible without the flexibility provided by natural language as a supervision signal and the domain-specific accuracy achieved through fine-tuning.

\section{Related Work}
\label{sec:related}

We divide relevant previous work into three main areas: model architecture, product representations, and general NLP challenges.

The model underlying this work, Contrastive Language–Image Pretraining (\texttt{CLIP})~\cite{Radford2021LearningTV}, builds on ideas from a rich literature involving zero-shot transfer, natural language supervision, and multimodal learning\footnote{When finalizing the submission, FAIR released \url{https://github.com/facebookresearch/SLIP}, which promises a slight improvement over \texttt{CLIP} for zero-shot classification: we leave the investigation of this modified contrastive architecture to future benchmarking.}. While our architecture re-purposes \texttt{CLIP}, an important capability we add through our domain adaptation is the ability to systematically vary input features.

Creating latent product representations is an active area of research, encompassing a variety of input data and downstream applications: content embeddings~\cite{wang-fu-2020-item}, behavioral embeddings  ~\cite{CasellesDupr2018Word2vecAT,bianchi2020fantastic} and hybrid embeddings~\cite{Vasile16}. Compared to previous work, our representations are multi-modal and transferable in zero-shot fashion to entirely new catalogs.

Finally, the literature on compositionality spans centuries: limiting ourselves only to recent work, \textit{grounding} has been explored in connection with efficient learning~\cite{Yu2018InteractiveGL,ChevalierBoisvert2019BabyAIAP}, and ``true understanding'' \cite{bender-koller-2020-climbing,Merrill2021ProvableLO}. Using combinatorial principles to test generalization abilities is a known strategy in toy worlds \cite{Chollet2019OnTM,Gandhi2021BabyIB}: we exploit insights from our target domain to operationalize similar principles on \textit{real-world} objects.

\section{Datasets and Model Training}
\label{sec:dataset}

\subsection{Training Dataset}
\label{sec:trainingdataset}

\textit{Farfetch} made available for the first time an English dataset comprising over 800k fashion products, with more than 3k brands across dozens of object types. Items are organized in hierarchical trees, producing a three-layer taxonomy: for example, \textit{trees} could be something like \textit{Clothing > Dresses > Day Dresses} or \textit{Clothing > Coats > Parkas}, for a total of 800+ trees. As input for the image encoder, we use the standard product image, which is a picture of the item over a white background, with no humans\footnote{Images follow a specific set of rules regarding the placement of the item, lights of the photo, etc., designed to highlight the item's features.}; as for the text, \textit{Farfetch} has two types of text, \textit{highlight} (e.g., ``stripes'', ``long sleeves'', ``Armani'') and \textit{short description} (``80s styled t-shirt'').

We create a training, validation and test set from the catalog by random sampling products. 
Our final training and validation sets comprise 700k and 50k products respectively from 188 categories.

\subsection{Testing Datasets}
\label{sec:testingdatasets}

Both for testing purposes and to further gauge the potential impact of the model in production at scale, we prepare the following datasets. \textbf{TEST} is the test set from \textit{Farfetch} containing 20k products; \textbf{HOUT-C} is the dataset containing a category which we excluded from training (\textit{Performance Tops}), for a total of 1.5k items; \textbf{HOUT-B} is the dataset containing two brands which were excluded from training, for a total of 1.7k items; \textbf{STLE} is a merchandising dataset from \textit{Farfetch}, completely independent from the catalog, that classifies 7749 items across 6 styles for gender women and 4 styles for gender men; example of styles are \textit{Classic} and \textit{Streetwear} and each item may belong to more than one style; \textbf{KAGL} is a subset of \citet{kaggle}, where each product has a white background image, a caption, and a category, for a total of 9990 items over 62 categories; \textbf{F-MNIST}~\cite{xiao2017/online} contains $10,000$ gray-scale images from 10 product classes, with pixel intensity inverted to obtain images with white background.\footnote{Note that these images have a size of 24x24 thus showing much less details than the images on which the models have been trained on.} \textbf{DEEP}~\cite{liuLQWTcvpr16DeepFashion} contains 4000 product images that are non-standardized (i.e contains humans) from 50 categories.\footnote{Authors and \textit{Farfetch} are working on releasing the dataset as well. Please check \url{https://github.com/Farfetch} for updates on the data release.}

\subsection{Training \texttt{FashionCLIP}}

We re-purpose the main architecture from~\cite{Radford2021LearningTV}, which we describe briefly here for the sake of completeness. \texttt{CLIP} is a multi-modal model comprising an image ($I_e(i_j)$) and a text ($T_e(t_k)$) encoder. During training, we sample $N$ pairs of $<i, t>$, and optimize a contrastive cross-entropy loss such that $I_e(i_j) \cdot T_e(t_k)$ for $j,k = 1,..,N$ is highest when the caption is paired with the correct image ($j=k$), and low otherwise ($j \neq k$). In the end, we obtain a multi-modal space where images and texts are jointly projected and learned: if training has been successful, we expect that, for example, the textual embedding for the string ``red long dress'' is actually similar (as measured by the dot product) to the image embeddings of red dresses. Table~\ref{tab:training} shows training time, performance and costs, and Appendix~\ref{sec:appendixTraining} contains full details to reproduce our setup. 


\begin{table}
\centering
\begin{tabular}{lcccc}
\hline
\textbf{LR}   & \textbf{Loss} & \textbf{Time(m)} & \textbf{USD} & \textbf{kgCO$_2$eq}\\
\hline
1e-4          & 16.0          & 618               & 31\$               & 0.77 \\
1e-5          & 1.73          & 617               & 31\$               & 0.77 \\
\textbf{1e-6} & \textbf{2.83} & \textbf{621}      & \textbf{31\$}      & \textbf{0.78}\\ \hline
\end{tabular}
\caption{Comparing training time, performance, costs and carbon emission on variants of the \texttt{FashionCLIP} architecture on the \textit{Farfetch} catalog. Cost is calculated with the AWS pricing for a \textit{p3.2xlarge}; estimations were conducted using the Machine Learning Impact calculator from \citet{lacoste2019quantifying}. Model used for testing in \textbf{bold}.}
\label{tab:training}
\end{table}

\section{Evaluation}
\label{sec:eval}
We benchmark \texttt{FashionCLIP} capabilities on several use cases, motivated by industry challenges, literature from \texttt{CLIP} and theoretical NLP considerations.

\subsection{Multi-modal Retrieval}
\label{sec:retrieval}

We test \texttt{FashionCLIP} on multi-modal retrieval to verify how pre-training may bring improvements for real-world product search. Our benchmark takes as input the caption of a product in the \textit{test set}, \textbf{TEST}, and asks models to rank product images corresponding to the caption -- the gold standard is the image associated with the product; in particular, \texttt{FashionCLIP} performs the dot product between the input caption embedding and each image vector embedding, obtained via $T_e$ and $I_e$ respectively and returns a rank based on descending order. We use ~\textit{HITS@k} as our metric.\footnote{In our experiments, $k$ is set to a challenging $5$.} Table~\ref{tab:retrieval} compares \texttt{FashionCLIP} against non-domain specific \texttt{CLIP}, and shows how fine-tuning significantly improves the understanding of our target domain.

\begin{table}
\centering
\begin{tabular}{llc}
\hline
\textbf{Model} & \textbf{Dataset} & \textbf{HITS@5}\\
\hline
\texttt{F-CLIP} & TEST & \textbf{0.61} \\
\texttt{CLIP} & & 0.22\\ 
\hline
\texttt{F-CLIP} & HOUT-C & \textbf{0.57} \\
\texttt{CLIP} &  & 0.28\\
\hline
\texttt{F-CLIP} & HOUT-B & \textbf{0.55} \\
\texttt{CLIP} &  & 0.27\\
\hline
\end{tabular}
\caption{Comparing \texttt{FashionCLIP} (F-CLIP) vs \texttt{CLIP} on the multi-modal retrieval task.}
\label{tab:retrieval}
\end{table}

\begin{figure}
  \centering
  \includegraphics[width=7.5cm]{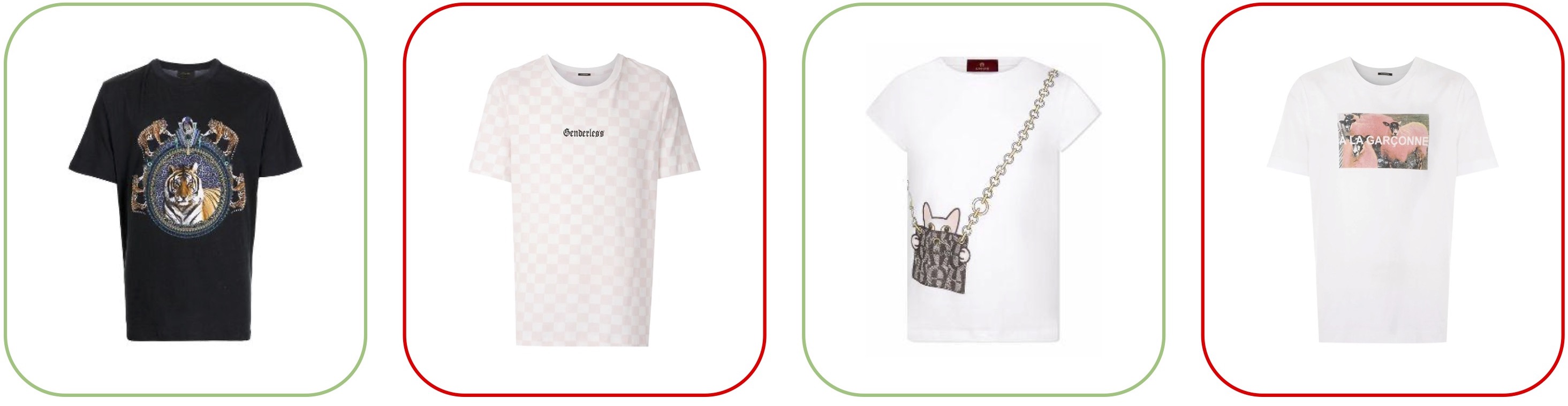}
  \caption{\textbf{Retrieval with non-fashion concepts}. Sample results for ``t-shirt with tiger'' and ``t-shirt with cat'' from \texttt{FashionCLIP} (\textit{green}) vs \textit{Farfetch}  production search engine (\textit{red}).}
  \label{fig:cats}
  \vspace{-4mm}
\end{figure}

We also perform extensive qualitative tests comparing \texttt{FashionCLIP} with the current search engine; Fig.~\ref{fig:cats} shows a case of particular interest for product search, that is, when visual concepts do not belong to the fashion domain and are not available in the caption. The first comparison (\textit{left}) shows that \texttt{FashionCLIP} is able to recover the concept of \textit{tiger} when prompted with ``t-shirt with tiger''; for the same query the search engine retrieves items matching the category, unable to interpret \textit{tiger} based solely on text. The second comparison (\textit{right}) shows that \texttt{FashionCLIP} is able to interpret \textit{cat} from a stylized, partially occluded drawing, whereas the search engine fails to generalize beyond the captions explicitly containing the string ``cat''. For more input-output pairs, please refer to the materials in Appendix~\ref{sec:appendixOutput}. 

\subsection{Zero-shot Classification}
\label{sec:classification}

The setup for the classification task follows~\citet{Radford2021LearningTV}: the model generates \textit{one} image embedding for the product image, and \textit{k} text embeddings, one for each of the labels in the classification scheme (e.g., ``shoes'', ``shirt''); the predicted label is the one that is most similar (through dot product) to the image. Table~\ref{tab:classification} summarizes the results of our benchmarks. On all the tested benchmarks, \texttt{FashionCLIP} is superior to \texttt{CLIP}, a result that suggests domain-specific fine-tuning is indeed useful in-domain and that it generalizes to other, completely unseen, datasets.


\begin{table}
\centering
\begin{tabular}{llccc}
\hline
\textbf{Model} & \textbf{Dataset} & \textbf{F1}\\
\hline
\texttt{F-CLIP} & TEST & \textbf{0.39} \\
\texttt{CLIP}   &      & 0.31\\
\hline
\texttt{F-CLIP} & KAGL  & \textbf{0.67} \\
\texttt{CLIP}   &       &   0.63\\
\hline
\texttt{F-CLIP} & F-MNIST & \textbf{0.71} \\
\texttt{CLIP}   &       &    0.66\\
\hline
\texttt{F-CLIP} & DEEP & \textbf{0.47} \\
\texttt{CLIP}   &      &    0.45    \\
\hline
\end{tabular}
\caption{Comparing the performance of \texttt{FashionCLIP} (F-CLIP) on product classification task over several datasets (\textbf{F1} is \textit{weighted macro F1}).}
\label{tab:classification}
\end{table}

Furthermore, we set out to investigate the ``cheating hypothesis'' on our domain-specific model, i.e. the hypothesis that supervised models do not generalize as well as \texttt{CLIP} because they fit to spurious features one dataset at a time. We freeze the image encoder from \texttt{FashionCLIP} and fine-tune a linear classifier, \texttt{LINEAR}, over the embeddings generated on a subset of categories (47) from the validation set from \textit{Farfetch}. We run benchmarks on $\mathbf{TEST_S}$, $\mathbf{KAGL_S}$, $\mathbf{F}$\textbf{-}$\mathbf{MNIST_S}$ and $\mathbf{DEEP_S}$, sub-sampled versions of respective datasets. Where labels are different, we adapt \texttt{LINEAR} to the labels by pooling the scores from relevant classes. We compare this to zero-shot performance, using the original labels to generate the text embeddings.

Table \ref{tab:cheating} reports our findings, which are  partially similar to those from \citet{Radford2021LearningTV}. Given that \textbf{F-MNIST} is very different from \textbf{TEST} \footnote{Comparable, for example, to CIFAR-100 \cite{Krizhevsky09learningmultiple} vs ImageNet \cite{deng2009imagenet}.}, the decrease in performance may be an indication of cheating. However, \texttt{LINEAR} performs well on the other datasets, with the biggest gain for \textbf{KAGL}, whose product images are the most resembling of those in \textbf{TEST} (i.e. high-resolution items on a white background). Compared to the original setting, it may be argued that the supervised model has an easier job in our case: fewer categories, and fairly homogeneous items, \textbf{F-MNIST} aside. 

\begin{table}
\centering
\begin{tabular}{lccc}
\hline
\textbf{Dataset} & \textbf{F-CLIP}  & \textbf{LINEAR} & \textbf{$\Delta$F1}\\
\hline
\texttt{$TEST_S$}   & 0.746 & 0.900 &\textbf{+0.154} \\
\texttt{$KAGL_S$}   & 0.764 & 0.881 & \textbf{+0.117} \\
\texttt{$DEEP_S$}   & 0.411 & 0.444 & \textbf{+0.033} \\
\texttt{$F$-$MNIST_S$}& 0.781 & 0.602 &\textbf{-0.179} \\
\hline 
\end{tabular}
\caption{\texttt{LINEAR} classification performance relative to zero-shot on \texttt{F-CLIP} (\textbf{F1} is \textit{weighted macro F1}).}
\label{tab:cheating}
\end{table}

While we leave as future work the investigation of fashion classification in more ecological settings, our results contain actionable insights for real-world deployments: in particular, supervised classifiers still require a good deal of manual intervention even for similar datasets, and they are utterly unusable on neighboring different problems. Table~\ref{tab:style-classification2} reports performance on \textbf{STLE}, with products still coming from \textit{Farfetch}, but where labels are manually assigned by merchandisers, and are completely orthogonal to the taxonomy (\textit{classic, streetwear, edgy} vs \textit{shoes, hats, bags}). The versatility afforded by language supervision allows zero-shot models to tackle the challenge by simple prompt engineering (``an item in \textit{classic} style''), while supervised models would require a new training and evaluation pipeline.

The trade-off between additional maintenance costs and accuracy gains is particularly important for B2B players (Section~\ref{sec:industry}): although there is no answer that can fit all the use cases, we wish to encourage a more data-driven decision process by charting the options and providing costs and performance assessment.

\begin{table}
\centering
\begin{tabular}{lcc}
\hline
\textbf{Model} & \textbf{Man} & \textbf{Woman} \\ \hline
\texttt{Prior} & 0.24 & 0.20 \\ \hline
\texttt{F-CLIP}  & \textbf{0.36} & \textbf{0.27} \\
\texttt{CLIP} & 0.33 & 0.17 \\ \hline
\end{tabular}
\caption{F1 macro on \textbf{STLE}; \texttt{Prior} classifies using empirical class probabilities.}
\label{tab:style-classification2}
\end{table}

\subsection{Compositionality}
\label{sec:grounding}

\begin{figure*}
  \centering
  \includegraphics[width=12cm]{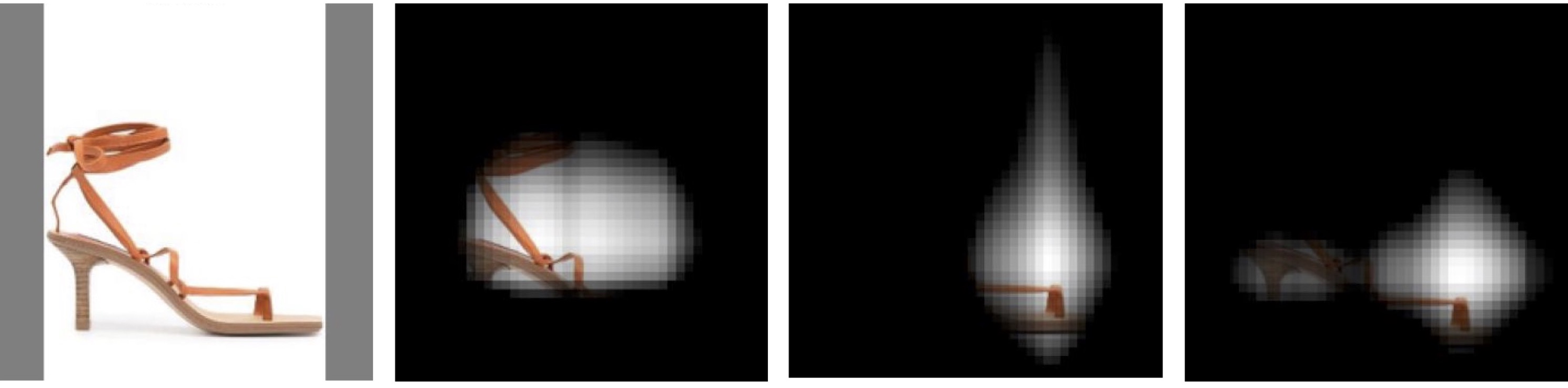}
  \caption{\textbf{Grounding and compositionality}. Localization maps for a product retrieved with the query ``ankle strap sandals with high heels'': left-to-right, the product, ``ankle strap'', ``sandals'', ``high heels'').}
  \label{fig:grounding}
  \vspace{-4mm}
\end{figure*}

Finally, we offer a preliminary investigation on \texttt{FashionCLIP} compositional abilities. Our analysis starts from two lessons from previous research. First, \textit{localization maps}~\cite{Fong_2017_ICCV,covert2021explaining} are an effective way to probe the model for \textit{referential} knowledge \cite{bianchi2021contrastive}\footnote{We borrow here the referential / inferential distinction from the classic~\citet{marconi}.}. Second, from a linguistic point of view most search queries in fashion have the form of Noun Phrases (NPs)  -- e.g. ``armani dress''. Therefore, the semantics of NP can be considered a good real-world generalization \cite{bianchi-etal-2021-language,bianchi-etal-2021-query2prod2vec}.

\begin{figure}
  \centering
  \includegraphics[width=7.5cm]{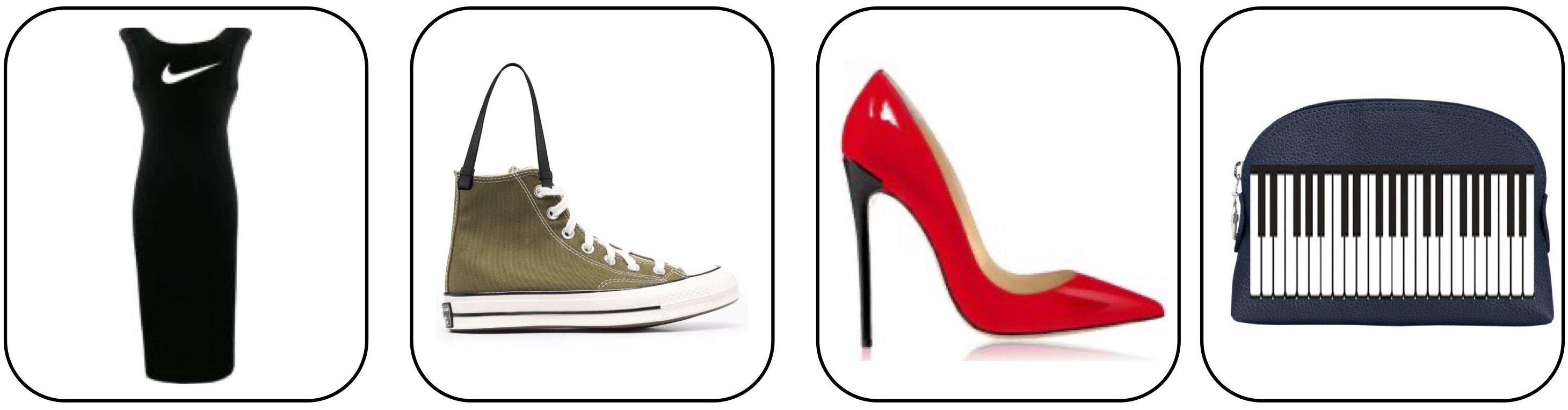}
  \caption{\textbf{Improbable products}. By combining fashion features, brands, items in new ways, we obtain visually realistic products with a clear, zero-shot compositional semantics. From left to right: ``Nike long dress'', ``converse with handles'', ``red shoes with black high heel'', ``keyboard pochette''.}.
  \label{fig:synth}
  \vspace{-4mm}
\end{figure}

Because isolated concepts reliably map onto visual regions, our working hypothesis is that \texttt{FashionCLIP} should exhibit true \textit{inferential} abilities in combining such concepts to generate new NPs. We build on domain knowledge, previous literature~\cite{liuLQWTcvpr16DeepFashion} and \textit{Farfetch}'s inventory to probe the model for knowledge of \textit{brands}  (e.g. ``nike''), \textit{features}  (``high heels'') and \textit{drawings} (``keyboard''). Since maps are easier to interpret than other probing strategies, we create localization maps and manually verify the text-to-region mapping (Appendix~\ref{sec:appendixMaps}) for each of these concepts. Since single concepts are grounded in regions (Fig.~\ref{fig:grounding}), the idea is to use that knowledge to \textit{systematically} generate new images and NPs. Crucially, we can assign a defined semantics to a new \textit{brand + object} NP that describes an ``improbable object'' that has never been seen before (Fig.~\ref{fig:synth})\footnote{Improbable objects vary: they may portrait odd combinations of concepts, such as a \textit{nike long dress}, surreal item, such as \textit{sneakers with handles}, or an unlikely extension of existing fashion items, such as the \textit{keyboard pochette} -- which generalizes the theme first found in J. Mugatu's \textit{keyboard tie}.}: a new NP such as ``nike dress'' would require the visual region corresponding to the word \textit{dress} to contain the visual region of the logo corresponding to the word \textit{nike}.

After verifying maps for improbable products through qualitative inspection, we supplement our analysis by re-purposing our classification and retrieval pipeline: in the classification task, \texttt{FashionCLIP} achieves an accuracy of $0.74$ when asked to pick the improbable label out of a set of distractors (Appendix~\ref{sec:appendixClassification}). For the retrieval task, we add the new images to \textbf{TEST}, and use the NPs as queries: out of 20k products, the model's top choice is correct half the time (\textit{Hit Rate@1}$=0.53$), a percentage that quickly rises to $0.82$ with $k=5$\footnote{As a comparison, \texttt{CLIP} scored \textit{Hit Rate@5}$=0.51$ and \textit{Hit Rate@1}$=0.73$}. 

While a full-fledged investigation of compositional abilities is beyond the scope of this contribution, \texttt{FashionCLIP} inferences on improbable products suggest the presence of \textit{some} degree of compositionality: important fashion concepts are ``identifiable'' in the latent space and can be singled out and re-combined into unseen concepts, exhibiting on a small scale the creative generalization we usually associate with symbolic systems~\cite{10.5555/335289}.

\section{Conclusion and future work}
\label{sec:conclusion}

We introduced \texttt{FashionCLIP}, a domain-adaption of \texttt{CLIP}, motivated by central use cases in fashion~\cite{10.1145/3447239}: differently from \textit{task-specific supervised} methods, \texttt{FashionCLIP} does not need a specialized architecture, labeling and tuning. We extensively verified the flexibility afforded by language supervision, and investigate semantic capabilities on new tasks. Our focus on a specific industry allows not just practical gains, but also opens up theoretical possibilities by constraining the domain, which is still large, but also easy to manipulate.

We only scratched the surface of what \texttt{FashionCLIP} can do: for example, multi-modal representations can be features in downstream systems, or directly used for zero-shot recommendations in item-to-item scenarios \cite{chia-etal-2022-does}; classification over arbitrary labels could be used as a fast and scalable labeling mechanism, supporting probabilistic labeling~\cite{Ratner2017SnorkelRT} or data generation for multi-modal IR models~\cite{10.1145/3366424.3386198}. While leaving this (and many other themes) to future iterations, we do believe \textit{this} work -- with its artifacts and methodology -- to be a first, rounded assessment of the great potential of \texttt{CLIP}-based models for e-commerce.


\section*{Ethical considerations}
The authors are aware of the risks of multi-modal \textit{CLIP}-like models in production associated with their limited robustness, as well as general issues with bias in large language models pre-trained at scale. In particular, we acknowledge that the risk of adversarial attacks on multi-modal models is an area of active research \cite{adversarial, 2005.10987}. To the limits of our knowledge, we have no reason to believe that \texttt{FashionCLIP} introduces any \textit{additional} risk when compared to the original CLIP. As with the original model, it should be noted that \texttt{FashionCLIP} appears to be susceptible to ``typographical attacks'' (Fig.~\ref{fig:apple}). No datasets used for training or testing contain PII and/or other sensitive user data.



\bibliography{anthology,custom}
\bibliographystyle{acl_natbib}

\appendix

\section{Training Routine and Parameters}
\label{sec:appendixTraining}
We apply fine-tuning starting from the pre-trained \texttt{CLIP} with the following parameters: we use Adam Optimizer with betas in (0.9, 0.98), epsilon of 1e-6 and weight decay equal to 0.2 and three different learning rates [1e-4, 1e-5, 1e-6]. We train the models for 4 epochs, evaluate every 500 steps and select the model with the lowest validation loss for each configuration (Table~\ref{tab:training}, model selected in \textbf{bold}). In our preliminary tests, the model with the lowest validation loss overall did not generalize the best in the zero-shot setting. This poses an interesting question, left for future work, of how to fine-tune these large pre-trained models without losing in generalization. The pipeline has been implemented with Metaflow \cite{Metaflow}, with training executed remotely on cloud GPUs; experiment tracking was provided by Comet \cite{CometML}.

\section{Input-Output Examples}
\label{sec:appendixOutput}

\begin{figure}[h]
  \centering
  \includegraphics[width=7.5cm]{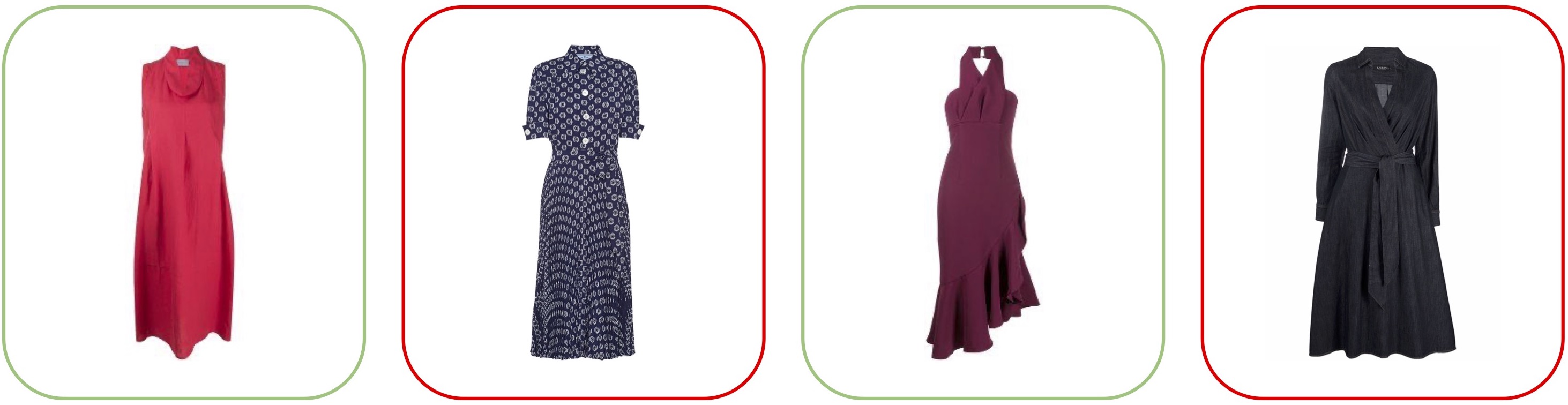}
  \caption{\textbf{\texttt{FashionCLIP} and color queries}. Retrieval results for ``light red dress'' and ``dark red dress'' from \texttt{FashionCLIP} (\textit{green}) vs \textit{Farfetch}  production search engine (\textit{red}).}
  \label{fig:red_dress}
  \vspace{-4mm}
\end{figure}

\texttt{FashionCLIP} multi-modal understanding is useful to resolve several difficult patterns in product search: non-fashion concepts (Fig.~\ref{fig:cats}), but also visual features that are either too subtle for pure text retrieval or ambiguous in the target domain.  Fig.~\ref{fig:red_dress} show a comparison for \textit{red}-themed queries, which combine both challenges: the distinction between lighter and darker shades of red is often not explicitly included in the product description, and ``red'' is a heavily ambiguous term (a color, but also the name of brands, product lines, etc.). Including multi-modal concepts such as those provided by \texttt{FashionCLIP} is a natural way to help a traditional search engine deal with these cases. 

\begin{figure}[h]
  \centering
  \includegraphics[width=5cm]{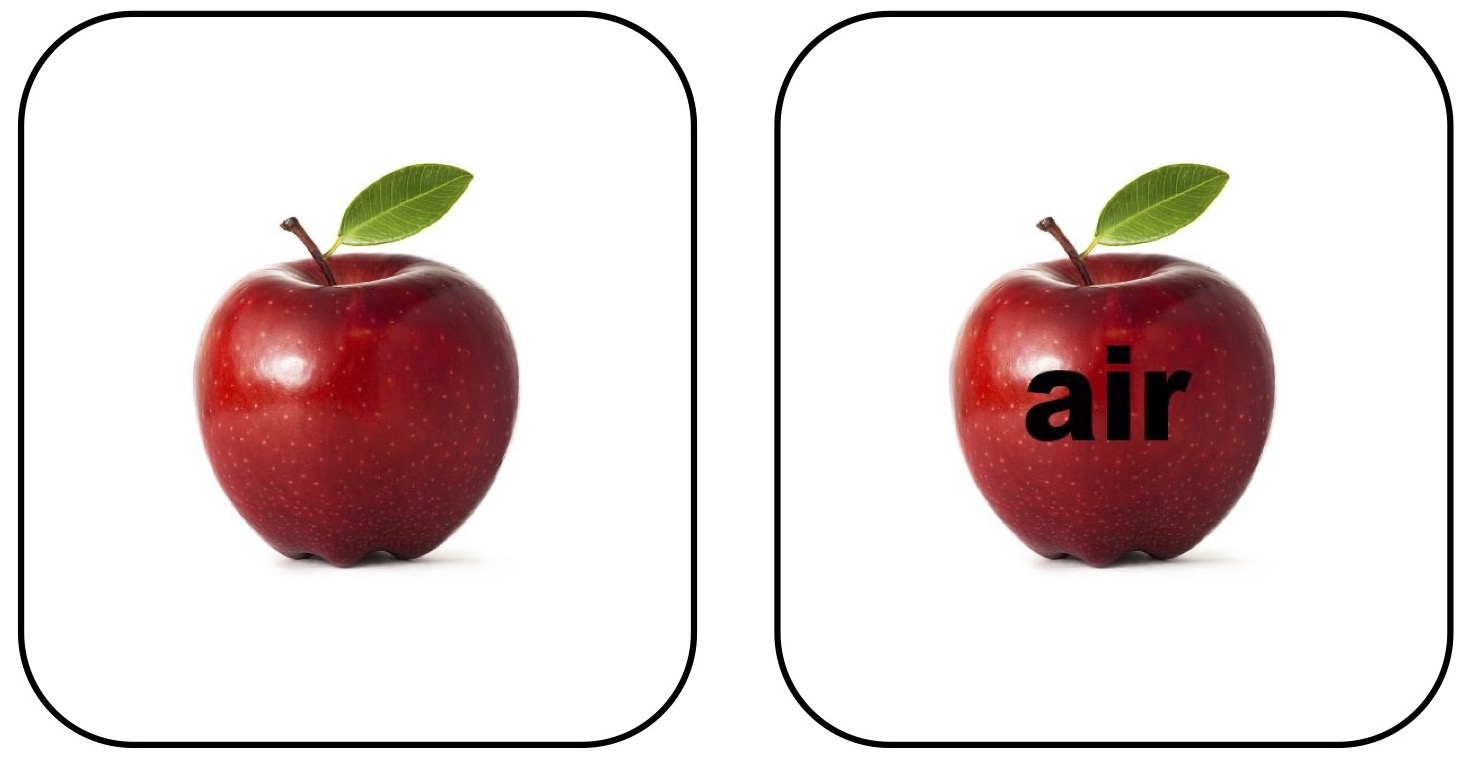}
  \caption{\textbf{Typographical attack}. \texttt{FashionCLIP} correctly identifies the object to the left as an ``apple'', but misclassifies the one to the right as ``nike air'', as the text acts as a confounder~\cite{verge}.}
  \label{fig:apple}
  \vspace{-4mm}
\end{figure}

Finally, since \texttt{FashionCLIP} learns a competent OCR (see also Appendix~\ref{sec:appendixGrounding}), it is not hard to fool the model into misclassifying items by simply printing out fashion-related words on them (Fig.~\ref{fig:apple}).

\section{Compositionality and Grounding}
\label{sec:appendixGrounding}

\subsection{Localization maps}
\label{sec:appendixMaps}

Section~\ref{sec:grounding} argues for the use of localization maps for a \textit{prima facie} verification of \texttt{FashionCLIP} referential abilities. Before constructing our improbable objects to test for \textit{inferential} abilities, we probe the model for visually grounded lexical knowledge. 

\begin{figure}[h]
  \centering
  \includegraphics[width=7.5cm]{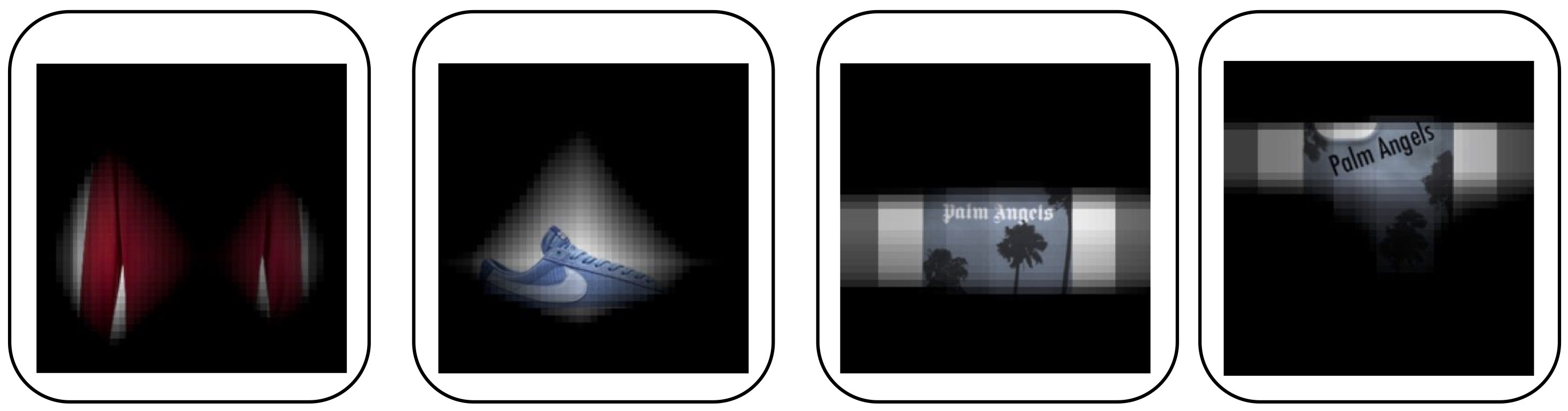}
  \caption{\textbf{Grounded lexical knowledge}. Maps are easy-to-use probes into the model fashion knowledge. \textit{Left to right}: localization map for ``long sleeves'' on a red polo; sneakers and the map for ``Nike'', a phone cover and the map for ``Palm Angels''; the same phone cover and map, when the logo is written with an out-of-distribution font in a new spot.}
  \label{fig:grounding2}
  \vspace{-4mm}
\end{figure}

We create localization maps by repeatedly occluding different parts of the image. We then encode each occluded version and measure its distance from the target text in the contrastive space. Intuitively, the farther the image is pushed away by the occlusion, the stronger was the linkage between the removed visual concept and the text and, in turn, the higher its score in the map. As shown in Fig.~\ref{fig:grounding} and \ref{fig:grounding2}, features such as ``high heels'', ``ankle strap'', ``long sleeves'' are well represented in \texttt{FashionCLIP}; the model seems also to be very aware of brands, in more or less explicit form: \texttt{FashionCLIP} picks up the abstract logo on \textit{sneakers} (Fig.~\ref{fig:grounding2}), as well as showing (similar to \texttt{CLIP}) good OCR capabilities, when recognizing a logo as an explicit text string.

\begin{figure}[t!]
  \centering
  \includegraphics[width=7.5cm]{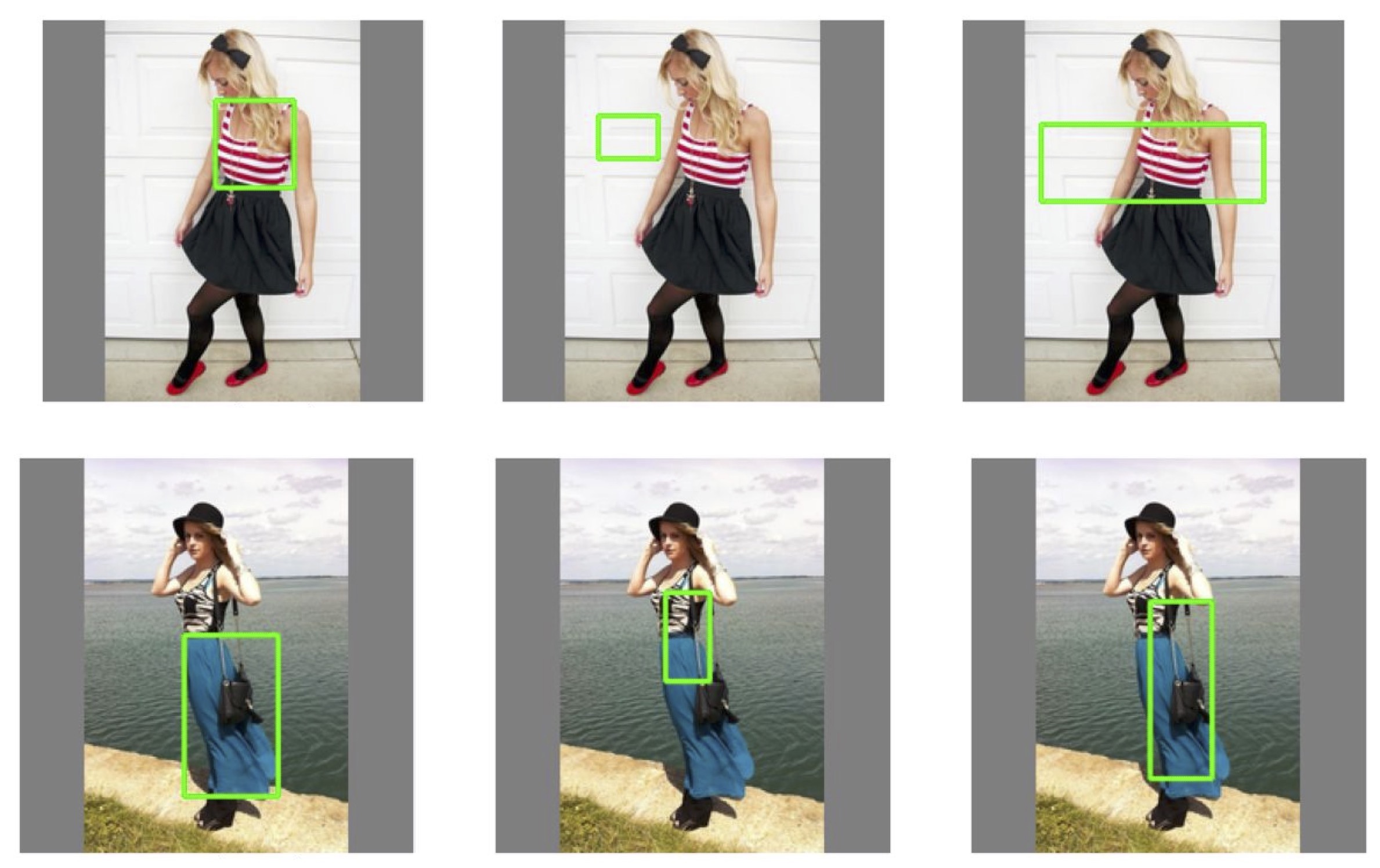}
   \caption{\textbf{Item segmentation}. Localization maps can be easily extended to provide zero-shot bounding boxes for item segmentation. \textit{Left to right}: ground truth, \texttt{CLIP}, \texttt{FashionCLIP} for ``top'' (top row), and ``skirt'' (bottom row).}
  \label{fig:segmentation}
  \vspace{-4mm}
\end{figure}

We also conducted preliminary experiments on \texttt{FashionCLIP} for zero-shot fashion parsing, which is an important open problem in the industry~\cite{10.1145/3447239}. The task is a specific case of semantic segmentation where bounding box annotations contain clothing items. We extract segmentation annotations from localization maps by finding the minimum bounding rectangle of highly activated areas.  Fig.~\ref{fig:segmentation} shows zero-shot annotations of some samples in the previously unseen ModaNet~\cite{zheng2018modanet} dataset. 

While it is unlikely that zero-shot models could replace specialized segmentation training, we believe that models such as \texttt{FashionCLIP} could find adoption as cheap, probabilistic labels for weak supervision pipelines, following our remarks in Section~\ref{sec:conclusion}.

\subsection{Classification for improbable products}
\label{sec:appendixClassification}

We analyze improbable products with zero-shot classification for preliminary evidence of inferential knowledge: in particular, we expect \texttt{FashionCLIP} to be able to assign the correct (improbable) label among credible distractors. The following are test examples:

\begin{itemize}
    \item \textbf{target}: \textit{NIKE DRESS} (as seen in Fig.~\ref{fig:synth}), \textbf{labels}: Nike dress, an Armani dress, a shirt, the flag of Italy, a Gucci dress,
    a Nike t-shirt;
    \item \textbf{target}: \textit{BLACK SHOES WITH RED HEEL}, \textbf{labels}: black shoes with red heel, black shoes, red shoes with red heel, red shoes with black heel, red shoes, fuchsia shoes, 
    the flag of Italy, sneakers, black sneakers, a bag.
    \item \textbf{target}: \textit{RED SHOES WITH BLACK HEEL} (as seen in Fig.~\ref{fig:synth}), \textbf{labels}: black shoes with red heel, black shoes, red shoes with red heel, red shoes with black heel, red shoes, fuchsia shoes, the flag of Italy, sneakers, black sneakers, a bag.
\end{itemize}

While exploring the full implications of this fact is outside the scope of \textit{this} work, distinguishing ``red shoes with black heel'' from ``black shoes with red hell'' implies knowledge beyond a bag-of-words semantics~\cite{Pham2021OutOO}.


\subsection{Sketch of a spatial semantics}
\label{sec:semantics}

As reported by \citet{bianchi-etal-2021-language}, NPs are a representative portion of the semantic variety to be found in fashion queries. To test for generalization in our multi-modal domain, we wish to \textit{systematically} connect (parts of) images with (parts of) queries: in other words, (parts of) images act as a \textit{domain} for the queries, in the model-theoretic sense \cite{10.5555/335289}. Consider the following language fragment:

\begin{enumerate}
    \item predicates are sorted in kinds: brands (e.g. \textit{nike(x)}), features (\textit{has\_heels(x)}), sortals (\textit{is\_dress(x)});
    \item a valid formula (representing a NP) is a combination of a sortal with any other type, e.g. \textit{is\_dress(x) + nike(x)}.
\end{enumerate}

Our domain is built out of pixels as atomic parts, so we may define \textit{regions} as a collection (i.e. a mereological sum) of pixels~\cite{10.5555/1213397}: given an image, an interpretation maps a predicate to regions (\textit{nike(x)} to the region occupied by the \textit{nike logo}). Depending on the kinds involved, we can proceed to define the NP semantics: a \textit{brand} region is a proper, internal part of a \textit{sortal} region, but a \textit{feature} may sometimes only be connected to a \textit{sortal} (e.g. ``converse with handles''). 

While all these relationships can be explicitly spelled out further (using for example mereo-topology~\cite{sep-mereology}), this draft hopefully provides an intuitive understanding of how we can systematically pair image editing and improbable labels, to produce the out-of-distribution products needed to test compositional inference.

\end{document}